\documentstyle[12pt]{article}
\begin{document}
\overfullrule=0pt

\addtolength{\baselineskip}{2mm}
\addtolength{\parskip}{1mm}


\newcommand{\hoi}{{\hbar \over i}}
\newcommand{\ah}{{1 \over 2}}

\newcommand{\tjr}{\tilde j^q}
\newcommand{\tcj}{ \tilde{\cal J}}
\newcommand{\tj}{ {\tilde J} }

\newcommand{\mf}{I}
\newcommand{\ms}{I\hskip -0.5mmI}
\newcommand{\mt}{I\hskip -0.5mmI\hskip -0.5mmI}

\newcommand{\by}{\bar y}
\newcommand{\bk}{\bar k}
\newcommand{\bya}{\bar y_1}
\newcommand{\byb}{\bar y_2}
\newcommand{\bza}{\bar k_1}
\newcommand{\bzb}{\bar k_2}
\newcommand{\pz}{\partial_k}
\newcommand{\pzb}{\partial_{\bar k}}

\newcommand{\te}{\tilde e}
\newcommand{\tm}{\tilde G^{\Delta}}

\begin{titlepage}
\begin{center}

{\Large \bf  The K-Z Equation and the Quantum-Group Difference Equation in}
{\Large \bf  Quantum Self-dual Yang-Mills Theory}

\vspace{3cm}

{\bf Ling-Lie Chau and Itaru Yamanaka} \\[1mm]

\vspace{5mm}

{\it Physics Department, University of California, Davis, CA 95616}
\end{center}

\vfill
\abstract{

{}From the time-independent current $\tcj(\bar y,\bar k)$ in the quantum
self-dual
Yang-Mills (SDYM) theory, we construct new group-valued
quantum fields $\tilde U(\bar y,\bar k)$ and $\bar U^{-1}(\bar y,\bar k)$
which satisfy a set of exchange algebras such that fields of $\tcj(\bar
y,\bar k)\sim\tilde U(\bar y,\bar k)~\partial\bar y~\tilde
U^{-1}(\bar y,\bar k)$ satisfy the original time-independent current
algebras.
 For the correlation
functions of the products of the $\tilde U(\bar y,\bar k)$ and $\tilde
U^{-1}(\bar y,\bar
k)$ fields defined in the invariant state constructed through the current
$\tcj(\bar y,\bar
k)$ we can derive the Knizhnik-Zamolodchikov (K-Z) equations with an
additional spatial dependence on $\bar k$.  From the $\tilde U(\bar y,\bar
k)$ and $\tilde U^{-1}(\bar y,\bar k)$ fields we construct the quantum-group
generators --- local, global, and semi-local --- and their algebraic
relations.
 For the correlation functions of the products of the $\tilde U$ and $\tilde
U^{-1}$
 fields defined in the
invariant state constructed through the semi-local quantum-group
generators we obtain the quantum-group difference equations. We give the
explicit solution to the two point function.
}

\vfill
\end{titlepage}

One of the pressing problems in particle physics is to formulate
four-dimensional (4-D) quantum
field theory nonperturbatively and to find nonperturbative
quantum-field-theoretical solutions to
the Yang-Mills equations for strong interactions, as well as to quantum
gravity. The path we have
taken in this pursuit of nonperturbative results has been through the
integrable-system method. The
main important framework the integrable-system provides is the possibility of
formulating the field
theory in terms of group-valued local fields.  This is a nontrivial starting
point for 4-D gauge
theory with nonvanishing curvatures. In this formulation the self-dual
Yang-Mill (SDYM) theory
is the simplest, yet important prototype to work out\cite{com1}.

In Ref.~\cite{cy3} we succeeded in formulating the quantum SDYM field theory
in
terms of the
group-valued local quantum field $\tilde J$, \cite{yang}. We  obtained the
interaction
Hamiltonian of the $\tilde
J$ fields, derived the exchange algebras that the $\tilde J$ fields satisfy,
showed that the
$\tilde J$ fields are bimodule quantum fields and the $R$ matrix of the
exchange algebras
satisfies the  Yang-Baxter relations so that the products of the $\tilde J$
fields satisfy
associativity, and we developed normal-ordering procedure for the products of
fields. From the
$\tilde J$ fields, we constructed local currents and their algebras. We found
that the
integrated currents in a particular spacial dimension and their current
algebras
are actually time-independent, i.e.
the currents commute with the
interaction Hamiltonian. This is a new feature in this 4-D quantum
field theory.
In Ref.~\cite{rtt}, we constructed  a time-independent local monodromy matrix
 $\tilde T \circ$
 from this time-independent currents
 and derived its
local exchange algebra
$R^T \tilde T \circ \tilde T \circ = \tilde T \circ \tilde T \circ R^T$ which
contains two infinitesimal forms, the time-independent infinite
local charge algebras and the infinite local Yangian algebras.
In this
paper we develop further the implications of these time-independent currents.

{}From the time-independent currents $\tcj(\bar y,\bar k)$, we construct new
group-valued
quantum fields $\tilde U(\bar y,\bar k)$ and $\bar U^{-1}(\bar y,\bar k)$
that satisfy a set of exchange algebras such that fields of $\tcj(\bar
y,\bar k)\sim\tilde U(\bar y,\bar k)~\partial\bar y~\tilde
U^{-1}(\bar y,\bar k)$ satisfy the original time-independent current
algebras.
(Through out the paper we use letters with superscript ``$~\tilde {~}~$" to
indicate
quantum operator fields.)
 For the correlation
functions of the products of the $\tilde U(\bar y,\bar k)$ and $\tilde
U^{-1}(\bar y,\bar
k)$ fields defined in the invariant state constructed through the current
$\tcj(\bar y,\bar
k)$ we can derive the Knizhnik-Zamolodchikov (K-Z) equations\cite{kz} with
an
additional spatial dependence on $\bar k$.  We can obtain the n-point
correlation functions of
the $\tilde U(\bar y,\bar  k)$ and $\tilde U^{-1}(\bar y,\bar k)$ fields;
they are expressible in
terms of the correlation functions of the quantum WZNW theory in
two-dimensions (2-D) with
coefficients being unknown functions of $\bar k$, one of the additional
spatial
coordinates in four dimensions.

  From the
$\tilde U(\bar y,\bar  k)$ and $\tilde U^{-1}(\bar y,\bar k)$ fields we can
also construct the
quantum-group  generators --- local, global, and semi-local ---
 and their algebraic
relations.
 For the correlation functions of the products of the $\tilde U$ and $\tilde
U^{-1}$
 fields defined in the
invariant state constructed through the semi-local quantum-group
generators, using the method given by Frenkel and Reshetikhin\cite{fr} we
obtain
the quantum-group difference equations. We give the  explicit solution to the
two point function.

With these results, we have exposed many important   quantum
integrability part
of this 4-D interactive theory. As expected, the 4-D interactive theory
is not, and should not be, as
fully integrable as integrable systems in 2-D. However it is to important to
find out the
quantum-field-theoretical integrability properties of the theories which have
many classical
integrability properties.  This experience in formulating the quantum SDYM
 from integralable-system point of view
 has now prepared us
to
investigate  fuller 4-D field theories.

{\it The Quantum SDYM System:  Hamiltonian, Exchange Algebras, Critical
Exponents, Normal-Ordering, and
Current Algebras }

First we briefly review the quantum SDYM theory formulated
in our previous
 paper\cite{cy3}. It is characterized by a
quantum field
Hamiltonian
\begin{equation}
\begin{array}{l}
\tilde H_{int} = - \alpha \int d{\bar y}~a^2 \Sigma_k\Sigma_{\bar k}
 \left\{
 Tr \{ (\pz \tilde J)(\pzb \tilde J^{-1})\} \right. \\
\\
\hspace{4.5cm} \left. + \int^1_0 d\rho Tr \{ (\partial_\rho \tilde {\tilde J}
)
[(\pzb \tilde {\tilde J}^{-1})(\pz \tilde {\tilde J})-
(\pz \tilde {\tilde J}^{-1})(\pzb \tilde {\tilde
J})](\tilde {\tilde J}^{-1})\}\right\},
\end{array}
\label{1}
\end{equation}
where, in the case of sl(2), $\tilde J = \tilde J (y,\bar y, k,\bar k)$
is a $2 \times 2$ matrix with non-commuting operator-valued entries
depending on the 4-d coordinates $ y, \bar y, k, \bar k; $ and $ y $ is the
 time. (Here we present the theory with $z$ and $\bar z$ coordinates
discretized: $z=ka$, $\bar z=\bar k a$, and $a\equiv l/N$ is the lattice
size.)
The field $\tilde {\tilde J}=
\tilde {\tilde J}(\rho; y,\bar y, k,\bar k)$ depends on a parameter $\rho$
and $ \tilde {\tilde J}(\rho=1)=\tilde J$.

The quantum
$\tj$ fields satisfy the following exchange algebras:

\begin{equation}
\begin{array}{l}
\tj_\mf(y,\bya,k_1,\bar k_1)\tj_{\ms}(y,\byb,k_2,\bar k_2)  \\
\hspace{1cm} =1_{\mf,\ms}\tj_{\ms}(y,\byb,k_2,\bar
k_2)\tj_\mf(y,\bya,k_1,\bar k_1)~
R_{\mf,\ms}\left(q,\bya-\byb\right)~~,
\end{array}
\label{2.a}
\end{equation}

\noindent where
\begin{equation}
R_{\mf,\ms}(q,\bya-\byb) =
P_{\mf,\ms}\left\{ [q]^{\triangle{_1}\ \varepsilon(\bya-\byb)}
{\cal P}_{j_{12} = 1}^q-[q]^{\triangle{_0}\ \varepsilon(\bya-\byb)}
{\cal P}^q_{j_{12}=0}\right\}~~,
\label{2.b}
\end{equation}

\begin{equation}
\varepsilon(\bya-\byb)
=-[ln(\bya-\byb+i\varepsilon)-ln(\byb-\bya+i\varepsilon)]/\pi i~~,
\label{2.c}
\end{equation}

\noindent and
\begin{equation}
\varepsilon(\bya-\byb)=\pm1,~ \quad{\rm for~~} \bya ~{}^{>}_{<}
{}~\byb~~;~~~
\label{2.d}
\end{equation}

\begin{equation}
\varepsilon(\bya-\byb)=0,~  \quad{\rm for~~} \bya = \byb~~;
\label{2.e}
\end{equation}
\begin{equation}
q\equiv e^{-[i\hbar/(4\alpha
{a^2})]\delta_{k_1k_2}\delta_{\bar k_1\bar k_2}},
\label{2.e1}
\end{equation}
 where $\alpha $ is the coefficient in front
of the SDYM interaction Hamiltonian, Eq.~(\ref{1});
$\triangle{_1}=2 {1 \over 2} ( {1 \over 2} +1 )-1(1+1) = -1/2$ and
$\triangle{_0}=2 {1 \over 2} ( {1 \over 2} +1 )-0(0+1) = 3/2$ are
differences of conformal dimensions of
two spin ${1 \over 2}$ fields minus that of a spin $j_{12}=1 $ and $0$
fields,
respectively
;
and the ${\cal P}^q_{j_{12}}$'s
are the $q$-ed projection matrices projecting
the two spin 1/2 states into $j_{12}=0$ or 1,
 satisfying ${\cal P}^q_{j_{12}}
{\cal P}^q_{j_{12}'} = {\cal P}
^q_{j_{12}} \delta_{j_{12}j_{12}'}$.
  In the more explicit expressions

\begin{equation}
{\cal P}^q_{j_{12}=1} = {\rm diag} \{1,d\left({q \atop 1}{1
\atop q^{-1}}\right), 1\}~~,
\label{2.f}
\end{equation}
where $d\equiv 1/(q+q^{-1})$.  The q-ed singlet projection matrix is
related to the triplet one by ${\cal P}^q_{j_{12}=0}=1-{\cal
P}^q_{j_{12}=1}$. The matrix $P_{\mf,\ms}$
 interchanges matrix in space I to II and vis versa, e.g.,
$P_{\mf,\ms}\tj_\mf (y,\bya,k_1,\bar k_1) \tj_{\ms}(y,\byb,k_2,\bar k_2)=
\tj_{\ms}(y,\bya,k_1,\bar k_1) \tj_\mf(y,\byb,k_2,\bar k_2) P_{\mf,\ms}$, and
its explicit
representation is
$P_{\mf,\ms}={1\over 2}+{1\over
2}\sum^3_{a=1}\sigma^a_\mf\sigma^a_{\ms}={\cal P}_{j_{12}=1} -
{\cal P}_{j_{12}=0}$; here the ${\cal P}_{j_{12}}$'s are the un-$q$-ed
ordinary projection matrices, i.e., Eq.~(\ref{2.f}) with $q=1$. The
subscripts I and II denote the
tensor spaces that the operator  matrices or c-number matrices operate on.
This tensor notation
saves us the
trouble of writing out the indices of the matrix elements;
 in terms of the matrix elements, Eq.~(\ref{2.a}) reads

\begin{eqnarray}
\lefteqn{
  \tj_{m_1, \alpha_1}(y,\bya,k_1,\bar k_1 )
\tj_{m_2,\alpha_2} (y,\byb,k_2,\bar k_2 )} \nonumber \\
&=&
\delta_{m_1,l_1}\delta_{m_2,l_2}
\tj_{l_2,\beta_2} (y,\byb,k_2,\bar k_2) \tj_{l_1,\beta_1}(y,\bya,k_1,\bar
k_1)
R_{\beta_1,\beta_2;\alpha_1,\alpha_2}(q,\bya-\byb).
\end{eqnarray}

 Using another
fact
$1_{\mf,\ms}= {\cal P}_{j_{12}=1} + {\cal P}_{j_{12}=0}$, we can
easily prove that at $\bya=\byb$, the exchange algebra Eq.~(\ref{2.a}) gives
\begin{equation}
{\cal P}_{j_{12}}\tj_\mf(y,\bya,k_1,\bar k_1)\tj_{\ms}(y,\bya,k_1,\bar
k_1)=\tj_\mf
(y,\bya,k_1,\bar k_1)\tj_{\ms}(y,\bya,k_1,\bar k_1){\cal P}^q_{j_{12}~~,}
\label{2.g}
\end{equation}
where $ j_{12}=0 , 1$. Eq.~(\ref{2.g})  implies ${\cal
P}_{j_{12}}\tj_\mf(y,\bya,k_1,\bar k_1)
\tj_{\ms}(y,\bya,k_1,\bar k_1)
 {\cal P}^q_{j'_{12}}=0$, for $j_{12}\not=j'_{12}$. This and the later
 development of the quantum group generators rely crucially on our
interpretation of the $R$ matrix at the coincidence
point, Eq.~(\ref{2.e}). We denote
$R_{\mf, \ms}(q,\bya-\byb)=R_{\mf, \ms}(q,+)$, for $\bya-\byb>0$ and
$R_{\mf, \ms} (q,\bya-\byb)=R_{\mf, \ms}(q,-)$, for $\bya-\byb<0$.  Note
that $[R_{\mf, \ms}(q,+)]^{-1} =R_{\ms, \mf}(q,-);~[R_{\mf,
\ms}(q,-)]^{-1}=R_{\ms, \mf}(q,+)$.

The expression for $\epsilon(\bya-\byb)$, Eq.~(\ref{2.c}), indicates that the
product $\tj_\mf (\bya) \tj_{\ms}(\byb)$ has singularity at
$\bya-\byb=0$ with the definite
 critical exponents given by
\begin{equation}
\begin{array}{l}
{\cal P}_{j_{12}}\tj_\mf(y,\bya,k_1,\bar k_1) \tj_{\ms}(y,\byb,k_1,\bar k_1)
{\cal P}^q_{j'_{12}} \\
\hspace{1.5cm}
=(\bya-\byb)^{ \triangle_{j_{12}}(ln \
q)/\pi i}\{:{\cal P}_{j_{12}}\tj_\mf(y,\bya,k_1,\bar k_1)\tj_{\ms}
(y,\byb,k_1,\bar k_1){\cal P}^q_{j'_{12}}:\}~~.
\end{array}
\label{3}
\end{equation}
(We call the power of the singularity $ { \triangle_{j_{12}}(ln \ q)/\pi i} $

the critical exponent of the
 product of the two $\tj $ fields.)
This also defines the normal-order products to be those in the curly
brackets;  their Taylor expansions give the operator-product expansions.

We then defined the $\tj^{-1}$ field by the following fixed-y-time equation

\begin{equation}
\tj(y,\bar y,k,\bar k) \tj^{-1}(y,\bar y,k,\bar k)=1
= \tj^{-1}(y,\bar y,k,\bar k)\tj(y,\bar y,k,\bar k)~~.
\label{4}
\end{equation}

\noindent From Eqs.~(\ref{4}) and (\ref{2.a}), we can easily show that the
$\tj^{-1}$
field
satisfies the following fixed-y-time exchange algebras

\begin{equation}
\begin{array}{l}
\tj^{-1}_\mf(y,\bya,k_1,\bar k
_1) \tj_{\ms}(y,\byb,k_2,\bar k
_2) \\
\hspace{1cm}
 =
\tj_{\ms}(y,\byb,k_2,\bzb)R^{-1}_{\mf,\ms}\left(q,\bya-\byb\right)
\tj^{-1}_\mf(y,\bya,k_1,\bar k
_1)~~,
\end{array}
\label{5.a}
\end{equation}

and
\begin{equation}
\begin{array}{l}
\tj^{-1}_\mf(y,\bya,k_1,\bar k_1) \tj^{-1}_{\ms}(y,\byb,k_2,\bar k
_2) \\
\hspace{1cm} =  R^{-1}_{\mf,\ms}\left( q,\bya-\byb\right)
\tj^{-1}_{\ms}(y,\byb,k_2,\bar k
_2) \tj^{-1}_\mf(y,\bya,k_1,\bar k
_1)~~.
\end{array}
\label{5.b}
\end{equation}

\noindent The construction of this $\tj^{-1}$ field is  crucial for us to
develop the full content of the theory in terms of the group-valued fields.

{}From $\hat H_{int}$ and the exchange algebra Eq.~(\ref{2.a}), we
derived
the equation of motion
\begin{equation}
\partial_y(\tj\partial_{\bar y}\tj^{-1})={\hbar\over
i}[\tilde H,\tj\partial_{\bar y}\tj^{-1}]
=a^{-2}\partial_k(\tj\partial_{\bar k}\tj^{-1})~~.
\label{5.c}
\end{equation}
{}From fields $\tj$ and
$\tj^{-1}$, we constructed
 the  $\widehat {{\rm sl}(2)}$ current
\begin{equation}
\tilde j(y,\bar y,k,\bar k) \equiv 2\alpha~\tj\partial_{\bar
y}~\tj^{-1}(y,\bar y,k,\bar k)~~.
\label{6}
\end{equation}
We then showed that the following equations can be easily derived from
the exchange algebras Eqs.~(\ref{2.a}),~(\ref{5.a}),~and~(\ref{5.b}),
\begin{eqnarray}
&&\hspace{-1cm}
{[\tilde j_\mf (y,\bya ,k_1,\bar k_1), \tilde j_{\ms}(y,\byb ,k_2,\bar
k_2)]}=
-i\hbar  [M_{\mf,\ms},~\tilde j_{\ms}(y,\byb ,k_2,\bar
k_2)] ~\delta(\bya -\byb)\delta_{k_1 k_2}\delta_{\bar k_1\bar
k_2}/a^2 \nonumber \\
&& -i\hbar 2\alpha ~M_{\mf,\ms}~\delta'(\bya -\byb)\delta_{k_1k_2}
\delta_{\bar k_1\bar k_2}/a^2 ~~,\label{7} \\
&&\hspace{-1cm}
[\tilde j_\mf (y,\bya ,k_1,\bar k_1),\tj_{\ms}(y,\byb ,k_2,\bar k_2)]
= -i\hbar M_{\mf,\ms}~\tj_{\ms}(\byb ,k_2,\bar k_2) ~\delta (\bya
-\byb)\delta_{k_1k_2}
\delta_{\bar k_1\bar k_2}/a^2~~, \label{8} \\
&&\hspace{-1cm}
[\tilde j_\mf(y,\bya ,k_1,\bar k_1), \tj^{-1}_{\ms}(y,\byb
,k_2,\bar k_2)]
= i\hbar  \tj^{-1}_{\ms}(\byb, k_2,\bar k_2)
M_{\mf,\ms}~\delta (\bya -\byb)
\delta_{k_1k_2}\delta_{\bza\bzb}/a^2, \label{9}
\end{eqnarray}
where $M_{\mf,\ms}\equiv P_{\mf,\ms}-{1\over 2}={1\over
2}\sum^3_{a=1}\sigma^a_\mf\sigma^a_{\ms}
\equiv -2 \sum^3_{a=1}\lambda^a_\mf\lambda^a_{\ms}$
and
$ [ \lambda^a , \lambda^b] = \varepsilon^{abc} \lambda^c$.
Eq.~(\ref{7}) is the current algebras of the currents $\tilde j$. Taking
trace of Eq.~(\ref{7}) onto $\lambda^a_\mf$ and $\lambda^b_{\ms}$ we can
easily obtain  current algebras in terms of
the Lie-components of the current
$[\tilde j^a(\bya),\tilde j^b(\byb)]=
i\hbar \varepsilon^{abc}\tilde j^c(\bya)~\delta(\bya-\byb)~
\delta_{k_1 k_2 }\delta_{\bar k_1\bar k_2}/a^2
- { i\hbar^2 \over 4\pi} K   \delta^{ab}~\delta'(\bya-\byb)
{}~\delta_{k_1 k_2 }\delta_{\bar k_1\bar k_2}/a^2 $,
where  \linebreak
$ K \equiv i \pi [a^2 ln(q)_{k_1=k_2,\bar k_1=\bar k_2}]^{-1}
= - 4 \pi \alpha / \hbar~$.
Eq.~(\ref{8}) indicates
that
  the left side of $\tj$  forms  the
fundamental representation of the
 current $\tilde j$; Eq.~(\ref{9}) indicates that the right side of
$\tj^{-1}$
forms
the fundamental representation of the current $\tilde j$. However, all these
relations are fixed
time relations because $\tilde j$ varies with time $y$, {\it i.e.},
$[\tilde j,\tilde H]\ne 0$.

Notice that we can easily obtain the continuum form of the current algebra
 by taking $ a \rightarrow 0$ and
$\delta_{k_1 k_2} /a \rightarrow  \delta ( z_1 - z_2 )$
and
$\delta_{\bar k_1\bar k_2}/a \rightarrow \delta( \bar z_1 - \bar z_2 ) $.

As pointed out in Ref.~\cite{cy3} the $k$-summed current
\begin{equation}
\tcj (\bar y,\bar k)\equiv a \Sigma_k\tilde j(\bar y,k,\bar k)~~,
\label{10}
\end{equation}
is constant in time $y$. It can be shown by directly calculating $[\tcj
,\tilde H_{int}]=0$ or
by summing the equation of motion in $k$
, Eq.~(\ref{5.c}). (Since $\tcj$ is time-independent
we do not
put $y$ dependence in $\tcj(\bar y,\bar k)$.)
Summing in $k$, notice that $\sum_{k_1}\delta_{k_1k_2}=1$,
from Eqs.~(\ref{7}) to (\ref{9}) we obtain the corresponding algebras,
\begin{eqnarray}
&& [\tcj_\mf(\bya ,\bza),\tcj_{\ms}(\byb,\bzb)] \nonumber \\
&& \hspace{-1cm}
=-i\hbar [M_{\mf ,\ms},\tcj_{\ms}
(\byb ,\bzb)]~\delta(\bya
-\byb)~\delta_{\bza\bzb}/a
-i\hbar 2 \alpha l
{}~M_{\mf ,\ms}~\delta' (\bya -
\byb) \delta_{\bza\bzb}/a ,
\label{11}
\end{eqnarray}
\begin{eqnarray}
&&[\tcj_\mf (\bya ,\bar k_1),\tj_{\ms}(\byb ,k_2,\bar k_2)]
= -i\hbar M_{\mf,\ms}~\tj_{\ms}(\byb ,k_2,\bar k_2) ~\delta (\bya -\byb)
\delta_{\bar k_1\bar k_2}/a ~~, \label{12} \\
&&[\tcj_\mf(\bya ,\bar k_1), \tj^{-1}_{\ms}(\byb
,k_2,\bar k_2)]
= i\hbar \tj^{-1}_{\ms}(\byb
,k_2,\bar k_2)~
M_{\mf,\ms}~\delta (\bya -\byb)
\delta_{\bza\bzb}/a~~,\label{13}
\end{eqnarray}
Taking trace of Eq.~(\ref{11}) onto $\lambda^a_\mf$ and $\lambda^b_{\ms}$
we can
easily obtain the current algebras in terms of
the Lie-components of the current
$
[\tcj^a(\bya),\tcj^b(\byb)]
=i\hbar \varepsilon^{abc}\tcj^c(\bya)~\delta(\bya-\byb)~\delta_
{\bza\bzb}/a  - { i\hbar^2 \over 4\pi} K'   \delta^{ab}~\delta'(\bya-\byb)
{}~\delta_{\bza\bzb}/a$,
where  $K'\equiv K N a= K l$.
 The corresponding continuum equations can also be
easily obtained.

{\it K-Z Equations for the Correlation Functions of the Products of
the Group-Valued Local Fields $\tilde U(\bar y,\bar k)$ and $\tilde
U^{-1}(\bar y,\bar k)$}

Next we can construct new time-independent group-valued local fields $\tilde
U(\bar y,\bar
k)$ and $\tilde U^{-1}(\bar y,\bar k)$ such that
\begin{equation}
\tcj_\mf(\bya ,\bza)= 2\alpha l~\tilde
U_\mf\partial_{\bya}~\tilde U^{-1}_\mf (\bya ,\bza)~~,
\label{14}
\end{equation}
\noindent and the $\tilde U$ and $\tilde U^{-1}$ fields satisfy the
following exchange algebras
\begin{equation}
\tilde U_\mf(\bya ,\bza)~\tilde U_{\ms}(\byb ,\bzb)= \tilde U
_{\ms}(\byb ,\bzb)~\tilde U_\mf(\bya ,\bza) ~R_{\mf ,\ms}\left(
q' ,\bya -\byb\right) ~~,
\label{15}
\end{equation}
\begin{equation}
\tilde U^{-1}_{\ms}(\byb ,\bzb)~\tilde U_{\mf}(\bya ,\bza)= \tilde U
_{\mf}(\bya ,\bza)~ R_{\mf ,\ms}\left(
q' ,\bya -\byb\right)~\tilde U^{-1}_{\ms}(\byb
,\bzb)~~,\label{16}
\end{equation}
\begin{equation}
\tilde U^{-1}_\mf(\bya ,\bza)~\tilde U^{-1}_{\ms}(\byb ,\bzb)=
{}~R_{\mf,\ms}
( q' ,\bya -\byb)~
  \tilde U^{-1}_{\ms}(\byb ,\bzb)~\tilde U^{-1}_\mf(\bya
,\bza)~~,\label{17}
\end{equation}
\noindent where ~~
\begin{equation}
q'\equiv  e^{[-i\hbar/(4\alpha l a)]\delta_{\bar k_1\bar k_2}}=
(q^{1/N})_{k_1=k_2}
\end{equation}
 and ~$ K'= i\pi [a~ ln(q')_{\bar k_1=\bar k_2}]^{-1}
$.
These exchange algebras guarantee the current algebras of $\tcj$,
Eq.~(\ref{11}),
and also gives the following algebra
\begin{equation}
[\tcj_\mf(\bya ,\bza),\tilde U_{\ms}(\byb ,\bzb)]=
-i\hbar  M_{\mf ,\ms}~\tilde U_{\ms}(\byb
,\bzb)~\delta(\bya-\byb)\delta_{\bza\bzb}/a~~,
\label{18}
\end{equation}
\begin{equation}
[\tcj_\mf(\bya ,\bza),[\tilde U_{\ms}(\byb ,\bzb)]^{-1}]=
i\hbar  [\tilde U_{\ms}(\byb
,\bzb)]^{-1} M_{\mf,\ms}~\delta(\bya-\byb)
\delta_{\bza\bzb}/a~~,
\label{19}
\end{equation}

\noindent and the algebra with $\hat U^{-1}(\bar y,\bar k)$.
{}From
\begin{equation}
2\pi
i~\delta(\bya-\byb)={1\over \bya-\byb-i\varepsilon}-{1\over
\bya-\byb+i\varepsilon},
\end{equation}
 the commutator equations
Eqs.~(\ref{7})~to~(\ref{9}) can be
written out in the commonly used operator-product-expansion forms
(which we leave as exercises for the reader).

Let us make the decomposition
\begin{equation}
\tcj(\bya ,\bza)=\tcj^+(\bya ,\bza)-\tcj^-(\bya
,\bza)~~,
\label{20}
\end{equation}

\noindent with $\tcj^\pm$ satisfying the following algebras
\begin{equation}
[\tcj^\pm_\mf(\bya ,\bza),~\tcj^\pm_{\ms}(\byb ,\bzb)]
= {-\hbar \over 2\pi} {1\over \bya -\byb+i\varepsilon}~[M_{\mf
,\ms},\left(\tcj^\pm_\mf\left(\bya
,\bza\right)+\tcj^\pm_{\ms}\left(\byb
,\bzb\right)\right)]\delta_{
\bza\bzb}/a ~~,
\label{21}
\end{equation}
\begin{eqnarray}
{[\tcj^+_\mf(\bya ,\bza),~\tcj^-_{\ms}(\byb ,\bzb)]}
&& \hspace{-5mm} = {-\hbar \over 2\pi} {1\over \bya -\byb-
i\varepsilon}~[M_{\mf
,\ms},\left(\tcj^+
_\mf\left(\bya ,\bza\right)+\tcj^-_{\ms}\left(\byb
,\bzb\right)\right)]\delta_{\bza\bzb}/a \nonumber \\
&& \hspace{1mm} - {\hbar \over 2\pi} {1\over (\bya -\byb-i\varepsilon)^2}~
2 \alpha l ~M_{\mf ,\ms}
\delta_{\bza\bzb}/a~~,
\label{22}
\end{eqnarray}

\begin{equation}
[\tcj^\pm_\mf(\bya ,\bza),\tilde U_{\ms}(\byb ,\bzb)]=
{-\hbar \over 2\pi} {1\over \bya -\byb\mp i\epsilon}~M_{\mf ,\ms}~\tilde
U_{\ms}(\byb
,\bzb)\delta_{\bza\bzb}/a~~,
\label{23}
\end{equation}

\noindent so that Eqs.~(\ref{11}), (\ref{18}), and (\ref{19}) are guaranteed.

Because of the singularities in the products of fields, we must prescribe
the normal-ordering procedure and make consistency checks.  The goal is
to obtain the following relation:
\begin{equation}
c~~\partial_{\bar y}\tilde U= :\tcj\tilde U :,
\label{23.b}
\end{equation}
 where
$  :\tcj\tilde U : =
-\tcj^+ \tilde U + (\tilde U^{T_{\ms}}
(\tcj^-)^{T_{\ms}} )^{T_{\ms}}$
and
$c$
is a constant to be determined by the normal-order procedure and
consistency. Following the procedure used in Ref.~\cite{fr}, we define the
vacuum state $|0\rangle$ to be
\begin{equation}
\tcj^- (\bar y,\bar
k)\mid0\rangle = 0~~,~~{\rm and~~} \langle0\mid\tcj^+ (\bar y,\bar
k)=0.\label{24}
\end{equation}
To determine the coefficient $c$, we take the commutator of
$\tcj^+$ and each side of Eq.~(\ref{23.b}), namely $\partial_{\bar y}\tilde
U$
 and $ :\tcj\tilde U :$. Here we consider the case $\bya \neq \byb$ and we
suppress $\varepsilon$.
\begin{equation}
\left[\tcj^+_\mf(\bya ,\bza),\partial_{\bar y_2} \tilde U_{\ms}(\byb ,\bzb)
\right]=
{-\hbar \over 2\pi} {1\over (\bya -\byb)^2}~M_{\mf ,\ms}~\tilde U_{\ms}(\byb
,\bzb)\delta_{\bza\bzb}/a~~,
\label{i1}
\end{equation}
\[
\left[\tcj^+_\mf(\bya ,\bza),~
-:\tcj_{\ms}(\bar y_2,\bar k_2)~\tilde U_{\ms}(\bar y_2,\bar
k_2): \right]
\]

\[
\equiv\left[\tcj^+_\mf(\bya ,\bza),
-\tcj^+_{\ms}(\bar y_2,\bar k_2)~\tilde U_{\ms}(\bar y_2,\bar
k_2)+\biggl(\Bigl(\tilde U_{\ms}(\bar y_2,\bar k_2)\Bigr)^{T_{\ms}}
{}~\Bigl(\tcj^-_{\ms}(\bar
y_2,\bar k_2)\Bigr)^{T_{\ms}}\biggr)^{T_{\ms}}\right]
\]

\begin{equation}
=
{-\hbar \over 2\pi} { 2 \alpha l \over (\bya -\byb)^2}~M_{\mf ,\ms}~\tilde
U_{\ms}(\byb
,\bzb)\delta_{\bza\bzb}/a
+ \left({\hbar \over 2\pi} \right)^2
 {2\over (\bya -\byb)^2}~M_{\mf ,\ms}~\tilde U_{\ms}(\byb
,\bzb)\delta_{\bza\bzb}/a^2
\label{i2}
\end{equation}

\begin{equation}
=
\left(
{-\hbar \over 2\pi}  2 \alpha l
+ \left({\hbar \over 2\pi} \right)^2
{2\over a}
\right)
{1\over (\bya -\byb)^2}~M_{\mf ,\ms}~\tilde U_{\ms}(\byb
,\bzb)\delta_{\bza\bzb}/a
\label{i2.2}
\end{equation}

\vskip 5mm
The consistency of $(\ref{23.b} )$ with Eqs. $(\ref{i1})$ and $(\ref{i2})$
requires
$c=(2 \alpha l - {\hbar \over 2\pi}{2\over a} )$ and we obtain
\begin{equation}
\bigl( 2 \alpha l - {\hbar \over 2\pi}{2\over a}
                   \bigr)~\partial_{\bar y}~\tilde U_\mf(\bar y,\bar
k)
  = -\tcj^+_\mf(\bar y,\bar k)~\tilde U_\mf(\bar y,\bar
k)+\biggl(\Bigl(\tilde U_\mf(\bar y,\bar k)\Bigr)^T~\Bigl(\tcj^-_\mf(\bar
y,\bar
k)\Bigr)^T\biggr)^T~~.
\label{25}
\end{equation}
\noindent Then using Eqs. (\ref{25}), (\ref{23}) and (\ref{24})
and following the procedure given in Ref.~\cite{kz},
we can straight forwardly
derive the K-Z equation for the $\tilde U$ fields,

\begin{equation}
\biggl(
-{1 \over 2}
\bigl(a~ K' + 2 \bigr)\partial_{\bar y{_j}}+
{1 \over 2}
\sum\limits_{k\ne
j}{M_{J,K}\over \bar y_j-\bar y_k}~\delta_{\bar
k{_j}\bar k{_k}}\biggr)\langle 0\mid \tilde U_\mf(\bya,\bza)\cdots \tilde
U_N(\bar y_n,\bar k_n)\mid 0\rangle=0~~,
\label{26.a}
\end{equation}
or
\begin{equation}
\biggl(
-{1 \over 2}
\bigl( a~K' + 2 \bigr)\partial_{\bar y{_j}} -
\sum\limits_{k\ne
j}{   \sum_a \lambda^a_J \lambda^a_K
\over \bar y_j-\bar y_k}~\delta_{\bar
k{_j}\bar k{_k}}\biggr)\langle 0\mid \tilde U_\mf(\bya,\bza)\cdots \tilde
U_N(\bar y_n,\bar k_n)\mid 0\rangle=0~~.
\label{26.b}
\end{equation}
Notice that taking away the $\bar k$ dependence, we recover the K-Z equation
of the quantum WZNW
theory.

The solutions of this SDYM K-Z equation are expressible in terms of those
of the WZNW K-Z equation, $\langle 0\mid g_\mf(\bya)\cdots g_N(\bar
y_n)\mid 0\rangle$, multiplied by unknown functions in $\bar k$.  For example

\begin{eqnarray}
& &\langle 0\mid \tilde U_\mf(\bya,\bza)\tilde U_{\ms}(\byb,\bzb)\mid
0\rangle = C_2(\bar k_1)\langle 0\mid \tilde g_\mf(\bya)\tilde
g_{\ms}(\byb)\mid
0\rangle_{(a K' + 2)}~\delta_{\bza\bzb}~~\nonumber \\
 & &+C_{21}(\bza,\bzb) \langle 0\mid \tilde g_\mf(\bya)\mid
0\rangle_{(a K' + 2)}  \langle 0 \mid \tilde g_{\ms}(\byb)\mid
0\rangle_{(a K' + 2)}~(1-\delta_{\bza\bzb}) ~
,\label{27} \\
 &  &\langle 0\mid \tilde U_\mf(\bya,\bar k_1)\tilde U_{\ms}(\byb,\bzb)
\tilde
U_{\mt}(\bar y_3,\bar k_3)\mid 0\rangle  \nonumber \\
&&\qquad= C_3(\bza)\langle 0\mid \tilde g_\mf(\bya)\tilde
g_{\ms}(\byb)\tilde g_{\mt}(\bar
y_3)\mid 0\rangle_{(a K' + 2)}~\delta_{\bza\bzb}\delta_{\bza\bar k_3}~~
\nonumber \\
&&+
\sum_{j\neq k \neq l}^{1 to 3}
 C_{3j}(\bar k_k , \bar k_j) \langle 0\mid \tilde g_J(\bar y_j) \mid
0\rangle_{(a K' + 2)}
\langle 0\mid \tilde g_K (\bar y_k)\tilde g_L(\bar
y_l) \mid 0\rangle_{(a K' + 2)}~\delta_{\bar k_k \bar k_l}(1-\delta_{\bar
k_k \bar k_j})~\nonumber \\
&&+
 C_{34}(\bza,\bzb,\bar k_3)
\langle 0\mid \tilde g_\mf(\bya)      \mid 0\rangle_{(a K' + 2)}
\langle 0\mid \tilde g_{\ms}(\byb)    \mid 0\rangle_{(a K' + 2)}
\langle 0\mid \tilde g_{\mt}(\bar y_3)\mid 0\rangle_{(a K' + 2)}\nonumber \\
&&~~~~~\times (1-\delta_{\bza\bzb})(1-\delta_{\bza\bar
k_3})(1-\delta_{\bzb\bar k_3})~~,\label{28}  \\
& & \nonumber \\
 &  &\langle 0\mid\tilde U_\mf(\bya,\bza)U_{\ms}(\byb,\bzb) U_{\mt}(\bar
y_3,\bar k_3) U_{IV}(\bar y_4,\bar k_4)\mid 0\rangle \nonumber \\
&&\qquad= C_4(k_1)\langle 0\mid\tilde g_\mf(\bya)\tilde g_{\ms}(\byb)\tilde
g_{\mt}(\bar
y_3)\tilde g_{IV}(\bar y_4)\mid
0\rangle_{(a K' + 2)}~\delta_{\bza\bzb}\delta_{\bzb,\bar
k_3}\delta_{\bar k_3\bar k_4} \nonumber \\
 &  & + \sum\limits^{1{\rm~to~}4}_{j\ne k\ne l\ne m}
\!\!\!
C_{4j}(\bar k_k,\bar k_j)
\langle 0\mid\tilde g_J(\bar y_j)\mid 0\rangle_{(a K' + 2)}
\langle 0\mid\tilde g_K(\bar y_k)
\tilde g_L(\bar y_l)\tilde g_M(\bar y_m)\mid 0\rangle_{(a K' + 2)}~\nonumber
\\
&&~~~~~\times(1-\delta_{\bar k{_j}\bar k{_k}}) \delta_{\bar k{_l}\bar k{_k}}
\delta_{\bar k{_m}\bar k{_k}}~\nonumber \\
  & & + \sum\limits^{1{\rm~to~}4}_{j\ne k\ne l\ne m}
\!\!\!C_{41,jklm}(\bar k_j,\bar k_l)\langle 0\mid\tilde g_J(\bar y_j)\tilde
g_K(\bar y_k)\mid
0\rangle_{(a K' + 2)}\langle 0\mid\tilde g_L(\bar y_l)\tilde g_M(\bar
y_m)\mid
0\rangle_{(a K' + 2)}~\nonumber \\
&&~~~~~\times \delta_{\bar
k{_j}\bar k{_k}}\delta_{\bar k{_l}\bar k{_m}}(1-\delta_{\bar k{_j}\bar
k{_l}})~\nonumber \\
  & & + \sum\limits^{1{\rm~to~}4}_{j\ne k\ne l\ne m}
\!\!\!C_{42,jklm}(\bar k_j,\bar k_k, \bar k_l)
\langle 0\mid\tilde g_J(\bar y_j)\mid 0\rangle_{(a K' + 2)}
\langle 0\mid \tilde g_K(\bar y_k)\mid 0\rangle_{(a K' + 2)} \nonumber \\
&&~~~~~\times \langle 0\mid\tilde g_L(\bar y_l)\tilde g_M(\bar y_m)\mid
0\rangle_{(a K' + 2)}~ (1-\delta_{\bar k{_j}\bar k{_l}})
\delta_{\bar k{_l}\bar k{_m}}
(1-\delta_{\bar k{_k}\bar k{_l}})(1-\delta_{\bar k{_j}\bar k{_k}})~ \nonumber
\\
  & & +
\!C_{43}(\bar k_1,\bar k_2, \bar k_3,\bar k_4)
\langle 0\mid\tilde g_\mf(\bar y_1)\mid 0\rangle_{(a K' + 2)}
\langle 0\mid \tilde g_{\ms}(\bar y_2)\mid 0\rangle_{(a K' + 2)} \nonumber \\
&&~~~~~\times \langle 0\mid\tilde g_{\mt}(\bar y_3)\mid 0\rangle_
{(a K' + 2)}
\langle 0\mid\tilde g_{IV}(\bar y_4)\mid 0\rangle_{(a K' + 2)}~\nonumber \\
&&~~~~~\times (1-\delta_{\bar k{_1}\bar k{_2}})
(1-\delta_{\bar k{_1}\bar k{_3}})
(1-\delta_{\bar k{_1}\bar k{_4}})
(1-\delta_{\bar k{_2}\bar k{_3}})
(1-\delta_{\bar k{_2}\bar k{_4}})
(1-\delta_{\bar k{_3}\bar k{_4}})
{}~~.
\label{29}
\end{eqnarray}
Taking away the $\bar k$-dependence
(and all $\delta_{\bar k{_i}\bar k{_j}}=1$)
, the correlation functions become
precisely those of quantum WZNW theory \cite{kz} .

These solutions indicate that we can obtain much information for this 4-D
 quantum field theory; however,
we can not obtain as much information as for
 the 2-D integrable system.
The quantum SDYM theory has
provided us a valuable example of how we can extract some exact and
nonperturbative information from the theory.

\pagebreak
{\it Quantum-Group Current $\tcj^q(\bar y,\bar k)$ and $\bar y$-Global
 Quantum-Group Generators $\tilde G(\bar k)$  }

 Similar to the construction of the current $\tcj$, Eq.~(\ref{14}), it is
natural
to construct the
other current,

\begin{equation}
 \tcj^q(\bar y,\bar k) \equiv 2\alpha l \tilde U^{-1}(\bar
y,\bar k) \partial_{\bar y}
\tilde U(\bar y,\bar k), \label{30}
\end{equation}
which we shall call
the quantum-group current since it has the quantum-group index on both
sides.

We can work out the algebraic relations among its matrix elements and with
the fields
$\tilde U$ and $\tilde U^{-1}$,  like
Eqs.~(\ref{11}) to (\ref{13})
for the $\tcj$. They all have nice quantum-group
interpretations.  However, we find that $\tcj^q$ is not as useful a quantity
as the
current $\tcj$ in that it can not be used to develop its vacuum states
and the corresponding differential equations as the current
$\tcj$ was used to develop the K-Z equations.
On the other hand we find that the following group-valued quantities, $\tilde
G(\bar k)$  and
$\tilde
G^{\Delta}(\bar y,\bar k)$, are the  appropriate quantum-group
generators for further development of the theory.

The $\bar y$-global quantum-group generator, denoted by
  $\tilde G(\bar k)$, is derived
from the
quantum-group current
$\tcj^q$ of Eq.~(\ref{30}) by a path-ordered
 integration,
\begin{equation}
\tilde G(\bar k) =\vec P exp\left(\int\nolimits^{\infty}_{-\infty} d\bar
y ~\tilde U^{-1}\partial_{\bar y}~\tilde U( \bar y,\bar k)\right)
=\tilde U^{-1}(\bar y=-\infty,\bar k)~\tilde U(\bar
y=+\infty,\bar k)~~.\label{31}
\end{equation}
Then, from the exchange algebras of the
fields $\tilde U$ and $\tilde U^{-1}$, Eqs.~(\ref{15}) to (\ref{17}), we can
derive the
algebraic relations among the
matrix elements of $\tilde G(\bar k)$ and with the fields
$\tilde U$ and $\tilde U^{-1}$,
\begin{eqnarray}
\lefteqn{ \hspace{-2cm} \{ R_{\ms , \mf}\left(q',+\right)~
\tilde
G_\mf(\bza) ~R_{\mf,\ms}\left(q',+ \right)\}
{}~\tilde G_{\ms}(\bzb)} \nonumber \\
&=&\tilde G_{\ms}(\bzb)~\{R_{\ms
,\mf}\left(q',+\right)~\tilde G_{\mf}(\bza)~
R_{\mf ,\ms}\left(q',+\right)\}~,\label{32}\\
\tilde G_\mf(\bza)~U_{\ms}(\byb,\bzb)
&=&U_{\ms}(\byb,\bzb)~\{ R_{\ms
,\mf}\left(q',+\right)~\tilde
G_\mf(\bza) ~R_{\mf
,\ms}\left(q',+\right)\}~,\label{33}\\
U^{-1}_{\ms}(\byb,\bzb)\tilde G_\mf(\bza)&=&\{ R_{\ms
,\mf}\left(q',+\right)~\tilde
G_\mf(\bar k_1)~R_{\mf,\ms}\left(q',+\right)\}~
\tilde U^{-1}_{\ms}(\byb,\bzb)~.\label{34}
\end{eqnarray}
where $R_{\mf ,\ms}(q',+)$ is the $R$-matrix with $\epsilon(\bya-\byb)=+1$.
In these equations, we use the curly bracket to bracket elements together to
make the content of the equation clearer.
 These three equations are the algebraic relations parallel to those
of
Eqs.~(\ref{11}),~(\ref{18}), and (\ref{19}). Associativity of all these
fields are true
because
the  $R$ matrix satisfies the
Yang-Baxter relations (which can be shown after some quite involved algebras.)

The basic elements of the quantum-group generators $\{\te_i(\bar k);i=3$
and $\pm\}$ are
related
to the components of the matrix
$\tilde G(\bar k)$ by
\begin{equation}
\tilde G(\bar k) \equiv
\left(   \begin{array}{cc}
                1 & 0 \\
               (1-q^2) ~\te_+(\bar k) & 1 \end{array} \right)
\left(   \begin{array}{cc}
               q^{-\te_3(\bar k)} & 0 \\
               0 & q^{\te_3(\bar k)}     \end{array}  \right)
\left(    \begin{array}{cc}
               1 & (q^{-1}-q) ~\te_-(\bar k) \\
               0 & 1                     \end{array}  \right),
\label{35}
\end{equation}
where the $\te_\pm(\bar k)$ and $q^{-{\te_3(\bar k)}}$ satisfy local
quantum-groups
algebras, which generalize those given in Ref.~\cite{fa1}.

{\it The $\bar y$-Semi-local Quantum-Group Generator $\tilde G^\Delta(\bar
y,\bar k)$}

Changing the integration range in Eq.~(\ref{31}) to a semi-local region we
obtain
the  $\bar y$-Semi-local Quantum-Group Generator $\tilde G^\Delta(\bar
y,\bar k)$ quantum group
generator:

\begin{eqnarray}
\tm(\bar y,\bar k)&\equiv &  \vec P exp \Biggl( \int\nolimits_{\bar
y-\Delta}^{\bar y+\Delta}d{\bar y}'~ \tilde U^{-1}(\bar
y',\bar k)~ \partial_{\bar y'}~
\tilde U(\bar y',\bar k)\Biggr) \nonumber\\
& =&\tilde U^{-1}(\bar y -\Delta,\bar k)~\tilde
U(\bar y +\Delta,\bar k)~~. \label{36}
\end{eqnarray}
We can easily show that $\tilde G^\Delta$ satisfies the
following algebras,
\begin{eqnarray}
\lefteqn{
 \{ R^{-1}_{\mf ,\ms}(q',\bar y_1-\bar
y_2)~\tm_\mf(\bya,\bza)~
 R_{\mf ,\ms}(q',\bya -\byb
+2\Delta)\}~\tm_{\ms}
(\byb,\bzb)} \nonumber \\
&=& \tm_{\ms}(\byb,\bzb)~ \{ R^{-1}_{\mf ,\ms}(q',\bya-\byb
-2\Delta)~\tm_\mf(\bya,\bza)~ R_{\mf ,
\ms}(q',\bya-\byb )\}~,\label{37} \\
\lefteqn{ \tm_\mf (\bya,\bza )~\tilde U_{\ms}(\byb,\bzb )} \nonumber \\
&=& \tilde
U_{\ms}(\byb,\bzb)~ \{ R^{-1}_{\mf ,\ms}
(q',\bya-\byb -\Delta)~
\tm_\mf(\bya,\bza)~ R_{\mf ,\ms}(q',\bya -\byb
+\Delta)\}~,\label{38}  \\
\lefteqn{
\tilde U^{-1}_{\ms}(\byb,\bzb)
 ~\tm_\mf(\bya,\bza)} \nonumber \\
&=&  \{ R^{-1}_{\mf ,\ms}(q',\bya
-\byb-\Delta)~\tm_\mf(\bya,\bza)~ R_{\mf
,\ms}
(q',\bya -\byb +\Delta)\} ~\tilde U^{-1}_{\ms}(\byb,\bzb)~.\label{39}
\end{eqnarray}

We next split the semi-local generator into the annihilation and creation
parts following a procedure similar to that used in Ref.~\cite{fr},

\begin{equation}
\tm_{\mf} (\bar y,\bar k) \equiv [\tilde G^{\Delta +}_{\mf}(\bar
y,\bar k )]^{-1} \tilde G^{\Delta
-}_{\mf}(\bar y,\bar k)~~,\label{40}
\end{equation}

\noindent and $\tilde G^{\Delta\pm}(\bar y,\bar k)$ satisfies the
following
 exchange algebras
\begin{eqnarray}
\lefteqn{
 R_{\mf ,\ms}(q',\bar y_1-\bar
y_2-\Delta)~ \tilde G^{\Delta\pm}_{\mf}(\bya,\bza)~
\tilde G^{\Delta\pm}
_{\ms}(\byb,\bzb) } \nonumber \\
&=& \tilde G^{\Delta\pm}_{\ms}(\byb,\bzb)~ \tilde G^{\Delta\pm}
_{\mf}(\bya,\bza)~ R_{\mf
,\ms}(q',\bya-\byb-\Delta)~, \label{41} \\
\lefteqn{
 R_{\mf ,\ms}(q',\bya-\byb)~ \tilde G^{\Delta
+}_{\mf}
(\bya,\bza)~ \tilde G^{\Delta -}_{\ms}(\byb,\bzb) } \nonumber \\
&=&  \tilde G^{\Delta -}_{\ms}(\byb,\bzb) \tilde G^{\Delta
+}_{\mf}(\bya,\bza)~ R_{\mf
,\ms}(q',\bya-\byb-2\Delta)~, \label{42}
\end{eqnarray}
\begin{equation}
\tilde U_{\mf}(\bya,\bza)~\tilde G^{\Delta\pm}_{\ms}(\byb,\bzb) =
\tilde G^{\Delta\pm}_{\ms}(\byb,\bzb)~ \tilde U_{\mf}(\bya,\bza)
{}~ R_{\mf ,\ms}(q',\bar y_1-\bar y_2 \pm
\Delta)~,\label{43}
\end{equation}

\noindent such that Eqs.~(\ref{37}) to (\ref{39}) are true.

Notice that
\begin{equation}
\Bigl[\sum_n \tcj(\bar y+n\Delta,\bar k),\tilde G^\Delta(\bar
y,\bar k)\Bigr]=0~~,\label{44}
\end{equation}
which manifests what we call the $\bar k$-local ${\rm
sl}^{\Delta}(n)\otimes U^{1/\Delta}q [{\rm sl}(n)]$
symmetry of the theory. For
$\Delta\rightarrow\infty$, Eq.~(\ref{44}) becomes $[\tcj(\bar y,\bar
k),\tilde
G(\bar k)]=0$, manifesting the $\bar k$-local $\widehat{{\rm
sl}(n)}\otimes Uq[{\rm sl}(n)]$ symmetry of the theory.
 For $\Delta\rightarrow 0$, Eq.~(\ref{44}) becomes
$[\tilde Q(\bar k),\tcj^q(\bar y,\bar k)]=0$,
where $\tilde Q(\bar k)\equiv
\int\limits^\infty_{-\infty}
\tcj(\bar y, \bar k)
d\bar y={\lim\atop{\Delta\to
0}}\sum^\infty_{n=-\infty}[\Delta
\tcj(\bar y+n\Delta,\bar k)]$ and
$\tcj^q(\bar y,\bar k)$
is from the coefficient of the $\Delta$-term in the
expansion of Eq.~(\ref{36}),
manifesting the $\bar k$-local
${\rm sl}(n)\otimes U_q^{\infty}[{\rm sl}(n)]$ symmetry of the theory.

{\it Quantum-Group Difference
Equation of the Correlation Functions Defined in the $\mid
0_q\rangle$-Vacuum}

Using Eq.~(\ref{40}), we rewrite Eq.~(\ref{36}) as
\begin{equation}
\tilde U(\bar y+\Delta,\bar k) = \tilde U(\bar y-\Delta,\bar k)
\tm(\bar y,\bar k)=\tilde U(\bar y -\Delta,\bar k)[\tilde G^{\Delta +}(\bar
y,\bar k)]^{-1}G^{\Delta
-}(\bar y,\bar k)~~.\label{45}
\end{equation}
Now we want to move $\Bigl(\tilde G^{\Delta +}(\bar y,\bar k)\Bigr)^{-1}$
to the left of $\tilde U(\bar y-\Delta,\bar k)$, since we shall consider the
vacuum expectation values of the $\tilde U$ fields by the vacuum $\mid
0_q\rangle$ defined by
\begin{equation}
\tilde G^{\Delta -}(\bar y,\bar k)\mid 0_q\rangle=\mid 0_q\rangle~,~~~{\rm
and}~~~\langle0_q\mid \tilde G^{\Delta +}(\bar y,\bar
k)=\langle0_q\mid~~.\label{46}
\end{equation}
To achieve that feat first we use Eq.~(\ref{43}) for $\bar y_1 =\bar y -
\Delta$
and $\bar y_2 = \bar y$ and interchange its r.h.s. and l.h.s.,
\begin{eqnarray}
\tilde G^{\Delta +}_{\ms}(\bar y,\bar k)~
\tilde U_{\mf}(\bar y -\Delta,\bar k)~
R_{\mf, \ms}(0)
&=&
\tilde U_{\mf}(\bar y -\Delta,\bar k)~
\tilde G^{\Delta +}_{\ms}(\bar y,\bar k).  \nonumber \\
\lefteqn{ \hspace{-6.5cm}
{\rm Multipling~ it~ by~ }
\left(\tilde G^{\Delta +}_{\ms}(\bar y,\bar k)\right)^{-1}
{\rm from~ both~ sides,~ we~ obtain} } \nonumber \\
\tilde U_{\mf}(\bar y -\Delta,\bar k)~
R_{\mf, \ms}(0)~
\left(\tilde G^{\Delta +}_{\ms}(\bar y,\bar k)\right)^{-1}
&=&
\left(\tilde G^{\Delta +}_{\ms}(\bar y,\bar k)\right)^{-1}~
\tilde U_{\mf}(\bar y -\Delta,\bar k).  \nonumber \\
\lefteqn{\hspace{-6.5cm}
{\rm Taking~ transpose~ in~ tensor~} \mf  {\rm ~space,~ we~ obtain}
}\nonumber \\
R_{\mf, \ms}^{T_{\mf}}(0)~
\tilde U_{\mf}^{T_{\mf}}(\bar y -\Delta,\bar k)~
\left(\tilde G^{\Delta +}_{\ms}(\bar y,\bar k)\right)^{-1}
&=&
\left(\tilde G^{\Delta +}_{\ms}(\bar y,\bar k)\right)^{-1}~
\tilde U_{\mf}^{T_{\mf}}(\bar y -\Delta,\bar k).   \nonumber \\
\lefteqn{\hspace{-6.5cm}
{\rm Multiplying~ it~ by~ }
\left(R_{\mf, \ms}^{T_{\mf}}(0)\right)^{-1}
{\rm ~from~ left,~ it~ becomes} }\nonumber \\
\tilde U_{\mf}^{T_\mf}(\bar y -\Delta,\bar k)~
\left(\tilde G^{\Delta +}_{\ms}(\bar y,\bar k)\right)^{-1}
&=&
\left(R_{\mf, \ms}^{T_{\mf}}(0)\right)^{-1}
\left(\tilde G^{\Delta +}_{\ms}(\bar y,\bar k)\right)^{-1}
\tilde U_{\mf}^{T_{\mf}}(\bar y -\Delta,\bar k).
\nonumber \\
\lefteqn{\hspace{-6.5cm}
{\rm Taking~ transpose~ in~ tensor~ } \mf  {\rm ~space,~ we~ obtain} }
\nonumber \\
\tilde U_{\mf}(\bar y -\Delta,\bar k)~
\left(\tilde G^{\Delta +}_{\ms}(\bar y,\bar k)\right)^{-1}
&=&
\left(
\left(R_{\mf, \ms}^{T_{\mf}}(0)\right)^{-1}
\left(\tilde G^{\Delta +}_{\ms}(\bar y,\bar k)\right)^{-1}
\tilde U_{\mf}^{T_{\mf}}(\bar y -\Delta,\bar k)
\right)^{T_{\mf}}.   \nonumber \\
\lefteqn{ \hspace{-6.5cm}
{\rm Taking~ trace~ in~ tensor~} {\ms} {\rm~ space,~ it~ becomes} }\nonumber
\\
\tilde U_{\mf}(\bar y -\Delta,\bar k)~
\left(\tilde G^{\Delta +}_{\mf}(\bar y,\bar k)\right)^{-1}
&=&
Tr_{\ms} \left[
\left(
\left(R_{\mf, \ms}^{T_{\mf}}(0)\right)^{-1}
\left(\tilde G^{\Delta +}_{\ms}(\bar y,\bar k)\right)^{-1}
\tilde U_{\mf}^{T_{\mf}}(\bar y -\Delta,\bar k)
\right)^{T_{\mf}}
P_{\mf, \ms}
\right]. \nonumber \\
\lefteqn{\hspace{-6.5cm}
{\rm Using~ the~ fact:~} (X^T)^T=X, {\rm ~we~ obtain}} \nonumber \\
&&
\hspace{-3.5cm}
=
Tr_{\ms} \left[
\left(
\left(\left(\tilde G^{\Delta +}_{\ms}(\bar y,\bar k)\right)^{-1}\right)
^{T_{\ms}}
\left(\left(R_{\mf, \ms}^{T_{\mf}}(0)\right)^{-1}\right)^{T_{\ms}}
\tilde U_{\mf}^{T_{\mf}}(\bar y -\Delta,\bar k)
\right)^{T_{\mf}T_{\ms}}
{}~P_{\mf, \ms}
\right].  \nonumber \\
\lefteqn{\hspace{-6.5cm}
{\rm Using~}
(P_{\mf, \ms})^{T_\mf T_{\ms}}=P_{\mf, \ms}, {\rm ~it~ becomes}}\nonumber \\
&&
\hspace{ -3.5cm}
=
Tr_{\ms} \left[
\left(
P_{\mf, \ms}
\left(\left(\tilde G^{\Delta +}_{\ms}(\bar y,\bar k)\right)^{-1}\right)
^{T_{\ms}}
\left(\left(R_{\mf, \ms}^{T_{\mf}}(0)\right)^{-1}\right)^{T_{\ms}}
\tilde U_{\mf}^{T_{\mf}}(\bar y -\Delta,\bar k)
\right)^{T_{\mf}T_{\ms}}
\right].  \nonumber  \\
\lefteqn{\hspace{-6.5cm}
 {\rm Moving~ } P_{\mf, \ms} {\rm ~inside,~we~ obtain} }\nonumber \\
&&
\hspace{ -3.5cm}
=
\left(
\left(
\left(\tilde G^{\Delta +}_{\mf}(\bar y,\bar k)\right)^{-1}
\right)^{T_{\mf}}
	\left(
	Tr_{\ms}~
	P_{\mf, \ms}
		\left(
        	\left(R_{\mf, \ms}^{T_{\mf}}(0)\right)^{-1}
                \right)^{T_{\ms}}
	\right)
\tilde U_{\mf}^{T_{\mf}}(\bar y -\Delta,\bar k)
\right)^{T_{\mf}}.
 \label{i4}
\end{eqnarray}
we finally reach
\begin{equation}
\tilde U(\bar y+\Delta,\bar k)=\left(\left(\left(\tilde G^{\Delta +}(\bar
y,\bar
k)\right)^{-1}\right)^T
\Upsilon ~\tilde U^T(\bar y-\Delta,\bar k)\right)^T \tilde G^{\Delta
-}(\bar y,\bar k)~~,\label{47}
\end{equation}

\noindent where the superscript $T$ means matrix transpose (
but the order of the operator stay the same)
;
$\Upsilon \equiv {q'+q'^{-1} \over q'^2+q'^{-2}}
\times diag(q',q'^{-1})$, which results from


\begin{equation}
\Upsilon_{\mf} = (Tr)_{\ms}\left(P_{\mf ,\ms}\left(\left(\left(R_{\mf
,\ms}(q',0)\right)^{T_\mf}\right)^{-1}\right)^{T_{\ms}}
\right)~~,\label{48}
\end{equation}
where the superscripts $T_\mf$ and $T_{\ms}$ indicate the transpose of
matrices in
the tensor spaces $I$ and $II$ respectively.

Using Eqs.~(\ref{32}) and (\ref{33}), we obtain the difference equation for
correlation function
\begin{eqnarray}
\lefteqn{
 \langle 0_q\mid \tilde U_\mf (\bya,\bza) \cdots \tilde U_L (\bar
y_l+2\Delta,\bar k_l) \cdots
\tilde U_N(\bar y_n,\bar k_n)
\mid 0_q\rangle} \nonumber \\
&=&\langle 0_q\mid \tilde U_\mf (\bar y_1,\bar k_1) \cdots
\tilde U_L (\bar y_l,\bar k_l)\cdots
\tilde U_N(\bar y_n,\bar k_n) \mid 0_q\rangle  \nonumber \\
&&
\hspace{-2cm}
\times R_{L,L-1}\left(q',\bar y_l-\bar
y_{l-1}\right)
\cdots R_{L,I}(q',\bar y_l-\bar y_1)~ \Upsilon_L
{}~R_{L,N}\left(q',\bar y_l-\bar y_n+2\Delta\right) \nonumber \\
&& \hspace{2.5cm}
\cdots R_{L,L+1}\left(q',\bar y_l-\bar
y_{l+1}+2\Delta\right)~~.        \label{49}
\end{eqnarray}
For the special case of $``+2\Delta$'' being at $\bar y_n$, Eq.~(\ref{49})
simplifies to the following cyclic relation
\begin{eqnarray}
\lefteqn{
\langle 0_q\mid \tilde U_\mf (\bar y_1,\bar k_1) \cdots \tilde U_N(\bar
y_n+2\Delta,\bar k_n) \mid
0_q\rangle} \nonumber \\
&=& \langle 0_q\mid \tilde U_N (\bar y_n,\bar k_n) \tilde U_\mf
(\bar y_1,\bar k_1)
\cdots \tilde U_{N-1}(\bar
y_{n-1},\bar k_{n-1}) \mid  0_q\rangle\Upsilon_N~~. \label{50}
\end{eqnarray}

\noindent For the two point function, Eq.~(\ref{49}) becomes
\begin{eqnarray}
\lefteqn{
\langle0_q\mid\tilde U_\mf(\bar
y_1,\bar k_1)\tilde U_{\ms}(\byb+2\Delta,\bar k_2)\mid0_q\rangle} \nonumber
\\
&=&\langle0_q\mid\tilde U_\mf(\bya,\bar k_1)
\tilde U_{\ms}(\bar y_2,\bar k_2)\mid0_q\rangle R_{\ms
,\mf}(q',\byb-\bya)\Upsilon_{\ms}~~.  \label{51}
\end{eqnarray}

\noindent Multiplying Eq.~(\ref{51}) from the right by ${\cal
P}^{q'}_{j_{12}=0}$
and using the fact $\langle0_q\mid\tilde U_\mf\tilde
U_{\ms}\mid0_q\rangle{\cal
P}^{q'}_{j_{12}=0}=\langle 0_q\mid\tilde U_\mf\tilde U_{\ms}\mid 0_q\rangle$,
which
can be shown using the definition of $\mid 0_q\rangle$ given by
Eq.~(\ref{46}),
whereby Eq.~(\ref{51}) becomes
\begin{eqnarray}
\lefteqn{
\langle 0_q\mid \tilde U_\mf (\bar y_1,\bar k_1) \tilde U_{\ms}(\bar
y_2+2\Delta,\bar k_2) \mid
0_q\rangle} \nonumber \\
&=& \langle 0_q\mid \tilde U_\mf (\bar y_1,\bar k_1) \tilde U_{\ms}(\bar
y_2,\bar k_2)
\mid 0_q\rangle~{q'}^{-\Delta_0\varepsilon
(\bar y_1-\bar y_2)}~
{q'+q'^{-1} \over q'^2+q'^{-2}}~~, \label{52}
\end{eqnarray}

\noindent where the last factor on the right is from $
{\cal P}^{q'}_{j_{12}=0}~R_{\mf
,\ms}(\bya-\byb)\Upsilon_{\ms}{\cal P}^{q'}_{j_{12}=0}={\cal
P}^{q'}_{j_{12}=0}
{q'}^{-{\Delta_0}\varepsilon(\bya-\byb)}(b/a)$ with $b/a\equiv
(q'+q'^{-1})/(q'^2+q'^{-2})=([2]_q')^2/[4]_q'$ and the fact that
${\cal P}^{q'}_{j_{12}=0}$ multiplying the vacuum expectation value becomes
unit.

Its solution can be easily found and written in the following form:
\[
\langle 0_q\mid\tilde U_\mf(\bya,\bza)\tilde U_{\ms}(\byb,\bzb)\mid
0_q\rangle
=\delta_{\bza,\bzb}
A_0(\bza) Exp \biggl\{ -\biggl({\bya-\byb\over
2\Delta}\biggr)ln\biggl({q'+q'^{-1}\over
q'^2+q'^{-2}}\biggr)+\biggl[\biggl({|\bya-\byb|\over
2\Delta}\biggr)
\]
\begin{equation}
+2\sum^\infty_{n=1}\theta\biggl(-{|\bya-\byb|\over
2\Delta}-n\biggr) \biggr] ln(q'^{\Delta_{0}})\biggr\}
+(1-\delta_{\bza,\bzb})A_1(\bza,\bzb)
,\label{53}
\end{equation}
\noindent where $A_0$ and $A_1$ are arbitrary functions; $\theta
(x)=0,{1\over
2},1$ for $x<0, x=0, x>0$, respectively. This expression for the
solution is continuous in the $\bya-\byb>0$ region.  For expressing
the solution in a function that is continuous the $\bya-\byb<0$ region,
we replace $\sum^\infty_{n=1}\rightarrow\sum^\infty_{n=0}$ in the square
bracket of
the above equations.

Taking away the $\bar k$-dependence
(and all $\delta_{\bar k{_i}\bar k{_j}}=1$)
, the correlation functions become
precisely those of quantum WZNW theory \cite{bim} .

\medskip
{\bf Acknowledgment}

One of us (LLC) would like to acknowledge the hospitality of the Theory Group
at CERN during the
Summer of 1995 where part of the work on this paper was carried out.

This work is supported in part by the U.S. Department of Energy (DOE).

\pagebreak

\vfill

\end{document}